\documentclass[fleqn]{elsart}
\usepackage{amsmath,graphicx,amssymb}
\newcommand{\uE}{\ensuremath{\mathrm{E}}}  
\newcommand{\ud}{\mathrm{d}}
\newcommand{\be}{\ensuremath{\frac{\mathrm{d} B (\mathrm{E}1)}
{\mathrm{d}E}}}                            
\newcommand{\urms}{\ensuremath{\mathrm{rms}}}
\newcommand{\ucm}{\ensuremath{\mathrm{cm}}}
\newcommand{\uc}{\ensuremath{\mathrm{c}}}
\begin{document}
\begin{frontmatter}
\title{Analytical E1 strength functions of
two--neutron halo nuclei: \nuc{11}{Li} and \nuc{14}{Be}}
\author[chalmers]{C.~Forss\'en\corauthref{cor}},
\corauth[cor]{Corresponding author.}
\ead{c.forssen@fy.chalmers.se}
\author[kurchatov]{V.~D.~Efros} and
\author[chalmers]{M.~V.~Zhukov}
\address[chalmers]{Department of Physics, Chalmers University of
Technology and G{\"o}teborg University, S--412~96 G{\"o}teborg, Sweden}
\address[kurchatov]{The Kurchatov Institute, 123182 Moscow, Russia}
\begin{abstract}
An analytical model, recently developed to study the electromagnetic
dissociation (EMD) energy spectra of two--neutron halo nuclei, is
applied to \nuc{11}{Li} and \nuc{14}{Be}. We find that a reliable set of
experimental data on EMD could help to resolve the problem concerning
the structure of the \nuc{11}{Li} ground state. For \nuc{14}{Be} we find
a mutual inconsistency between the existing experimental data on the
binding energy, radius and EMD energy spectrum. We also conclude that
the structure of \nuc{14}{Be} is essentially more complicated than the
structure of other Borromean two--neutron halo nuclei.
\end{abstract}
\end{frontmatter}
\section{Introduction}
%
The reaction of Coulomb excitation in collisions with heavy targets is
an important tool to study the structure of unstable nuclei.  For nuclei
having only one bound state the excitation means breakup and the process
is called electromagnetic dissociation (EMD). In the present paper E1
EMD of two--neutron halo nuclei is considered. Microscopic calculations
of the process at the three--cluster level were performed for
\nuc{6}{He}~\cite{dan98:632,cob98:58},
\nuc{11}{Li}~\cite{cob98:58,tho98:24}, and
\nuc{14}{Be}~\cite{des95:52,tho96:53}. Such calculations are hindered by
incomplete knowledge of the cluster dynamics. Quantum Monte Carlo
A--nucleon calculations~\cite{pan99:654} are hindered by the complexity
of the A--body problem. Therefore it seems to be of interest to describe
the process semi--phenomenologically aiming both to understand its main
features and to guide experimental studies. Such an approach was
developed in our recent paper~\cite{for02:697}. Three--cluster
bound--state wave functions (WFs) of two--neutron halo nuclei were
constructed which behave correctly at large intercluster distances,
reproduce the nuclear sizes, and incorporate the main features of the
underlying three--body structure. No--interaction three--body WFs were
used to describe the breakup final states. The model allows an analytic
calculation of the strength functions and can serve as a tool to predict
the Coulomb disintegration spectra of a variety of two--neutron halo
nuclei. It may be used when precise knowledge concerning the structure
of a system under study is lacking. As a test case the relatively
well--studied nucleus \nuc{6}{He} was considered in~\cite{for02:697}. We
found that the large--distance asymptotics of the ground--state WF
determines the shape and the position of the maximum of the strength
function, while the size of the ground state governs the asymptotic
constant and thus the magnitude of the strength function. As to the
final--state interaction (FSI) effects, one can note that the total E1
strength will not change when replacing a complete set of true final
states with a complete set of no--interaction final states. Therefore
the inclusion of FSI will result only in a redistribution of the
strength leading to a somewhat higher strength at low energies. But even
without including FSI a very good agreement concerning the shape and
peak position of the E1 strength function and a reasonable agreement
concerning its magnitude was found in the \nuc{6}{He}
case~\cite{for02:697}.

In the present paper we apply the model to study the Borromean halo
nuclei \nuc{11}{Li} and \nuc{14}{Be}. For both nuclei there are large
ambiguities concerning the ground--state structure. Furthermore, there
are large uncertainties in the experimental data regarding EMD reactions
and, for \nuc{14}{Be}, also concerning basic ground--state quantities
such as binding energy and size. Therefore we believe that an analytical
model with large freedom in trying different assumptions can be very
useful.

In Sec.~\ref{sec:model} the model is outlined, more details can be found
in Ref.~\cite{for02:697}. In Sec.~\ref{sec:results} the results of our
analysis of the \nuc{11}{Li} and \nuc{14}{Be} EMD data are
presented. Sec.~\ref{sec:discussion} contains the conclusions.
%
\section{Analytical E1 strength functions of two--neutron halo nuclei%
\label{sec:model}}
%
We consider the electromagnetic excitation of a system with only one
bound state and consequently with all possible final states belonging to
the continuum. In the framework of first--order perturbation theory the
energy spectrum for E1 Coulomb excitation can be written as
\begin{equation}
  \frac{\d \sigma(\uE 1)}{\d E} = \frac{N_{\uE 1}(E^*)}{\hbar
  c}\frac{16 \pi^3}{9} \be.
\label{eq:xsec}
\end{equation}
Here $E^*$ is the excitation energy, $E^*=E_{0}+E$, where $E_{0}$ is the
binding energy, and $E$ is the continuum energy, $N_{\uE 1} (E^*)$ is
the spectrum of virtual photons~\cite{win79:319,ber85:442} and $\d B(
\uE1 )/ \ud E$ is the dipole strength function
\begin{equation}
\be=\frac{1}{2J_i + 1}
  \sum_{M_i}\sum_{\mu=-1,0,1}\int \d \tau_f\left| \langle f|
 \mathcal{M}(\uE 1,\mu) | i; J_i M_i \rangle \right|^2
  \delta\left(E_f-E\right),
\label{eq:strengthfuncdef}
\end{equation}
where $\d \tau_f$ is the phase--space element for final states,
$\vec{\mathcal{M}}(\uE 1)$ is the dipole transition operator,
$|i\rangle$ is the initial state, and $|f\rangle$ are the final states
in the center--of--mass (CM) system. The notation $\int \d \tau_f$ in
Eq.~\eqref{eq:strengthfuncdef} implies summation over discrete quantum
numbers in addition to integration.  We will study halo nuclei which are
known to exhibit a large degree of clusterization. Since we are
interested in low--energy excitations, involving mainly relative motion
between the clusters, we will use the corresponding $N$--cluster E1
operator
\begin{equation}
  \vec{\mathcal{M}}(\uE 1) = \sqrt{\frac{3}{4\pi}}
  \sum_{i=1}^N e Z_i (\vec{r}_i - \vec{R}_\ucm),
\label{eq:reldipoleop}
\end{equation}
where $\vec{r_i}$ are the cluster positions, and $\vec{R}_\ucm$ is the
position of the CM of the system.

 We adopt the three--body model to describe two--neutron halo
nuclei. The cluster part of the bound--state WF, in the CM system, is
written as an expansion over hyperspherical harmonics (HH), see
e.g.~\cite{zhu93:231},
\begin{equation}
\Psi\left( \vec{x},\vec{y} \right) = \rho^{-5/2}
\sum_{KLSl_xl_y} \chi_{KLS}^{l_xl_y} \left( \rho \right)
\left[ \Gamma_{KL}^{l_xl_y} \left( \Omega_5 \right) \otimes \theta_S
\right]_{JM}.
\label{eq:hhwf}
\end{equation}
Here $\{\vec{x},\vec{y}\}$ is the set of normalized Jacobi vectors where
$\vec{x}$ corresponds to the vector joining the two valence neutrons,
and $\{\rho,\Omega_5\}$ are the hyperspherical coordinates in the
$\{\vec{x},\vec{y}\}$ space.  The functions $\Gamma_{KLM}^{l_xl_y}$ are
HH, and $\theta_{SM_S}$ ($S=0,1$) are the spin functions of the two
valence neutrons, Since the HH expansion converges rapidly we retain
only one, or a few, terms. Furthermore, since the hyperradial functions,
$\chi_{KLS}^{l_xl_y} \left( \rho \right)$, should behave rather
similarly at large $\rho$, which is the region of interest, we use the
same hyperradial function for all terms retained in the expansion. These
approximations lead to the following normalized WF for initial bound
states having $J^\pi = 0^+$
\begin{equation}
\begin{split}
\Psi(\vec{x},\vec{y}) = \frac{\chi^{(N)} (\rho)}{\rho^{5/2}}
\left\{ \left[ a_{00} \Gamma_{000}^{00} \left( \Omega_5 \right) +
a_{20} \Gamma_{200}^{00} \left( \Omega_5 \right)
\right. \right. \\
\left. \left. + a_{40} \Gamma_{400}^{00} \left( \Omega_5 \right)
\right] \theta_{00} 
+ a_{21} \left[ \Gamma_{21}^{11} \left( \Omega_5 \right)
\otimes \theta_1 \right]_{J=0} \right\}.
\label{eq:modelwf}
\end{split}
\end{equation}
The coefficients $a_{00}$, $a_{20}$, $a_{21}$ and $a_{40}$ entering
here are the amplitudes of various HH, $a_{00}^2 + a_{20}^2 + a_{21}^2 +
a_{40}^2=1$.  (In addition to the WF components used
in~\cite{for02:697}, the $K=4$ contribution is introduced here and in
Eq.~\eqref{eq:str} below.) We consider hyperradial functions of the
following two forms. The first is a simple normalized exponential
\begin{equation}
\chi^{(1)} (\rho) \equiv \sqrt{2\kappa_0} \exp(-\kappa_0 \rho),
\label{eq:1pwf}
\end{equation}
where the single free parameter $\kappa_0$ is fitted to the binding
energy via $E_0 = (\hbar \kappa_0)^2 /( 2 m)$, and $m$ is the nucleon
mass. This function reproduces the true asymptotic behaviour of
Borromean three-body systems with two neutrons as constituents, see
e.g.~\cite{mer74:19}.  However, the three--body size $\langle \rho^2
\rangle$, and consequently the total size of the system, is
underestimated with such a model.  Therefore, we introduce a second
hyperradial model function having two free parameters
\begin{equation}
\begin{split}
\chi^{(2)} (\rho) & \equiv c\left[ \exp(-\kappa_0 \rho) -
\exp(-\kappa_1 \rho) \right], \\ 
& \mathrm{where} \quad c= \sqrt{\frac{2 \kappa_0 \kappa_1
(\kappa_0 + \kappa_1)} {(\kappa_0 - \kappa_1)^2}}.
\label{eq:2pwf}
\end{split}
\end{equation}
The parameters $\kappa_0$ and $\kappa_1$ are fixed using experimental
values of binding energy and rms radius, see~\cite{for02:697}. The
condition that $\kappa_1 > \kappa_0$ ensures that the second term
decays faster than the first, and thus the correct large--$\rho$
asymptotics is preserved.

We use a complete set of no--interaction final states that include
just coordinate--space HH
\begin{equation}
\frac{J_{K+2}(\kappa\rho)}{(\kappa\rho)^2}
\left[ \Gamma_{KL}^{l_xl_y} \left( \Omega_5 \right) \otimes \theta_S
\right]_{JM}.
\label{eq:cwf}  
\end{equation}
Alike the plane waves, these states are solutions to the free--space 
six--dimensional Schr{\"o}dinger equation. The continuum energy is
related to $\kappa$ via  $E =( \hbar \kappa)^2 /( 2 m)$.

Using these model WFs an analytical calculation of the transition matrix
elements (MEs) entering Eq.~\eqref{eq:strengthfuncdef}
is possible. When the model WF of Eqs.~(\ref{eq:modelwf},
\ref{eq:2pwf}) is adopted for the bound state, one obtains the following
final expression for the E1 strength function:
\begin{multline}
\be=c^2 D E^3\sum_{i,j=0}^1
\frac{(-1)^{i+j}}{\left[ (E_i+E) (E_j+E) \right]^{11/4}} \\
\times \left[\alpha_1 F_1(y_i)F_1(y_j)+
\alpha_2F_2(y_i)F_2(y_j) + \alpha_3F_3(y_i)F_3(y_j) \right],
\label{eq:str}
\end{multline}
Here $E_{0,1}=(\hbar\kappa_{0,1})^2/(2m)$, where $\kappa_0$,
$\kappa_1$ and $c$ are defined in Eq.~\eqref{eq:2pwf}. The constant
$D$ is
\begin{displaymath}
D=\frac{3}{2} \left( \frac{\hbar^2}{2m}
    \right)^{3/2} \frac{(Z_\uc e)^2}{(A- 2)A},
\end{displaymath}
and the constants $\alpha_1$, $\alpha_2$ and $\alpha_3$ are   
\begin{align*}
  \alpha_1 &= \frac{1}{4} \left( \frac{315}{2^{10}} \right)^2 \left(
  4a_{00}^2 + a_{20}^2 + a_{21}^2 + 4a_{00}a_{20} \right), \\
  \alpha_2 &= \left( \frac{9009 \sqrt{3}}{2^{17}} \right)^2 \left( 
  a_{20}^2 + a_{21}^2 + \frac{4}{9}a_{40}^2 + \frac{4}{3}a_{20}a_{40}
  \right), \\
  \alpha_3 &= \frac{3}{2} \left( \frac{36465}{2^{20}} \right)^2 a_{40}^2.
\end{align*}
The coefficients $a_{00}$, $a_{20}$, $a_{21}$ and $a_{40}$ 
are defined in Eq.~\eqref{eq:modelwf}. The
quantities $y_{0,1}$ are defined as $y_i=E/(E+E_i)$, and, finally,
\begin{align*}
F_1(y) &= F\left( \frac{11}{4}, \frac{3}{4}, 4,y\right),\qquad
F_2(y)  = y F \left( \frac{15}{4}, \frac{7}{4}, 6,y\right), \\
F_3(y) &= y^2 F \left( \frac{19}{4}, \frac{11}{4}, 8,y\right),
\end{align*}
where $F(\alpha, \beta, \gamma, z)$ is the standard hypergeometrical
function. The corresponding three terms in Eq.~\eqref{eq:str}
represent contributions from final states with $K=1$, $K=3$ and $K=5$
respectively. When the hyperradial function~\eqref{eq:1pwf} is used
the result is obviously obtained from Eq.~\eqref{eq:str} by retaining
only the term with $i=j=0$ and replacing $c^2$ with $2\kappa_0$.
%
\section{Results%
\label{sec:results}}
\subsection{\nuc{11}{Li}}
Despite the great number of experimental and theoretical studies that
have been performed, the \nuc{11}{Li} nucleus still remains a puzzle.
The role of possible $s$--wave intruder
states~\cite{joh90:244,tho94:49,uet99:59,suz00:662,gar02:700} is of much
theoretical interest. The experimental evidence for virtual $s$--states
in the binary subsystem \nuc{10}{Li} is diverse and sometimes
conflicting~%
\cite{wil75:59,ame90:52,you94:49,zin95:75,boh97:616,%
tho99:59,cag99:60,cha01:510}.
Three--body microscopic calculations of \nuc{11}{Li} are uncertain due
to this fact, in particular.  Note that whereas the most recent
experiments establish that the ground state of \nuc{10}{Li} is a
$s_{1/2}$ state~\cite{cha01:510}, this itself does not mean that the
$(s_{1/2})^2$ configuration is a predominant one in \nuc{11}{Li}. The
weights of the $(s_{1/2})^2$ and $(p_{1/2})^2$ configurations depend on
the relative strength of $s$-- and $p$--wave potentials determining the
gap between the $s_{1/2}$ and $p_{1/2}$ states in \nuc{10}{Li}. The
shape of the \nuc{10}{Li} momentum distribution, measured in a
\nuc{11}{Li} fragmentation reaction~\cite{sim99:83}, supports a
$(1s_{1/2})^2$ contribution of $(45 \pm 10 )$\% in the \nuc{11}{Li}
ground--state WF. Further evidence for the mixing of different parity
states is clearly seen in the asymmetric angular
correlations~\cite{sim99:83}.

Clearly, the present--time understanding of the structure of
\nuc{11}{Li} is substantially poorer than that in the \nuc{6}{He} case,
in which case more definite information on the interaction between the
valence neutrons and the core is available. This makes \nuc{11}{Li} an
interesting nucleus to study with our analytical model. In this way we
will be able to test different assumptions on the ground--state
structure. In the present work we have used four different ground--state
WFs of \nuc{11}{Li}, which are summarized in Table~\ref{tab:wfs}. Here
we restricted ourselves to HH with the lowest possible $K$ values, $K=0$
and $K=2$. These configurations correspond, respectively, to the
$(s_{1/2})^2$ and $(p_{1/2})^2$ configurations of the core--centered
shell model for valence nucleons.

To explain the latter fact we need to consider a relationship between HH
and the configurations of the core--centered shell model. One can pass
from the Jacobi vectors $\{\vec{x},\vec{y}\}$ defined in connection with
Eq.~\eqref{eq:hhwf} to another pair $\{\vec{x}_1,\vec{y}_1\}$ of Jacobi
vectors, such that $\vec{x}_1$ is proportional to the distance between
one of the valence nucleons and the core, and $\vec{y}_1$ is
proportional to the distance between the other valence nucleon and the
CM of the pair consisting of the former valence nucleon and the
core. The distance from the CM of this pair to the core is $A_c^{-1}$
times smaller then that to the valence nucleon belonging to the pair. In
the framework of the approximation where one neglects the former
distance as compared to the latter one, i.e. locates the CM of the pair
on the core, the Jacobi orbital momenta associated with the vectors
$\{\vec{x}_1\}$ and $\{\vec{y}_1\}$ coincide with orbital momenta of the
two valence nucleons in the core--centered shell model.  In order to
find HH to be retained in the model WF we proceed as follows. Let
$\Gamma_{KLM}^{l_xl_y}(\vec{x},\vec{y})$ denote the HH entering
Eq.~\eqref{eq:hhwf}. In addition, we define the HH
$\Gamma_{KLM}^{l_{x_1}l_{y_1}}(\vec{x}_1,\vec{y}_1)$ where $l_{x_1}$ and
$l_{y_1}$ are Jacobi orbital momenta associated with $\vec{x}_1$ and
$\vec{y}_1$. We consider first the expansion over the latter HH and we
retain the terms in this expansion for which the orbital momenta $L$,
$l_{x_1}$, and $l_{y_1}$ are the same as for predominant configurations
in the core--centered shell model. Note, however, that the corresponding
HH do not coincide with the lowest shell--model configurations but also
include higher shell--model states with radial excitations.  The final
formula~\eqref{eq:modelwf} with the coefficients listed in
Table~\ref{tab:wfs} is obtained expressing the retained HH
$\Gamma_{KLM}^{l_{x_1}l_{y_1}}(\vec{x}_1,\vec{y}_1)$ as linear
combinations of the HH $\Gamma_{KLM}^{l_xl_y}(\vec{x},\vec{y})$. The
quantum numbers $K$ and $L$ are conserved at this transformation of HH.

The first two of our WFs are 100\% $K=0$ states. In the case of $K=0$
states the orbital momenta of relative motion in each pair of the three
particles are zeros. Thus the $K=0$ state corresponds to the
$(s_{1/2})^2$ configuration of the core--centered shell model. The wave
function $\Psi_1$(\nuc{11}{Li}) contains the one--parameter hyperradial
function~\eqref{eq:1pwf} while $\Psi_2$(\nuc{11}{Li}) contains the
two--parameter function~\eqref{eq:2pwf}.

The two other WFs, $\Psi_3$(\nuc{11}{Li}) and $\Psi_4$(\nuc{11}{Li}),
include HH with $K=0$ and $K=2$. They correspond to 50\% mixtures of
$(s_{1/2})^2$ and $(p_{1/2})^2$ configurations. To obtain the $K=2$ HH
and their weights listed in Table~\ref{tab:wfs} we expand the
$(p_{1/2})^2$, $J=0$ neutron shell--model state over such states in $LS$
coupling. In the framework of our infinite core--mass approximation the
latter states correspond to HH, $\Gamma_{KLM}^{l_1l_2}$, with $K=2$ and
$l_1=l_2=1$, defined with respect to the above mentioned Jacobi vectors
$\{\vec{x}_1, \vec{y}_1\}$. Two such HH arise, one with $L=1$, and the
other with $L=0$. Finally, when the above mentioned rotation to the
$\{\vec{x}, \vec{y}\}$ system is performed the first HH (with $L = 1$)
transforms into a single HH with $K=2, \; L=l_x=l_y=1$ and the second
one (with $L = 0$) transforms into a single HH with $K=2, \;
L=l_x=l_y=0$.

The WF $\Psi_4$(\nuc{11}{Li}) differs from $\Psi_3$(\nuc{11}{Li}) only
in that the relative signs of the amplitudes pertaining to the
$(p_{1/2})^2$ configuration and the $(s_{1/2})^2$ configuration are
different.

\begin{figure}[hbt]
  \begin{center}
  \begin{minipage}{0.95\textwidth}
    \begin{minipage}[t]{0.47\textwidth}
      \centering
      \includegraphics[width=\textwidth]{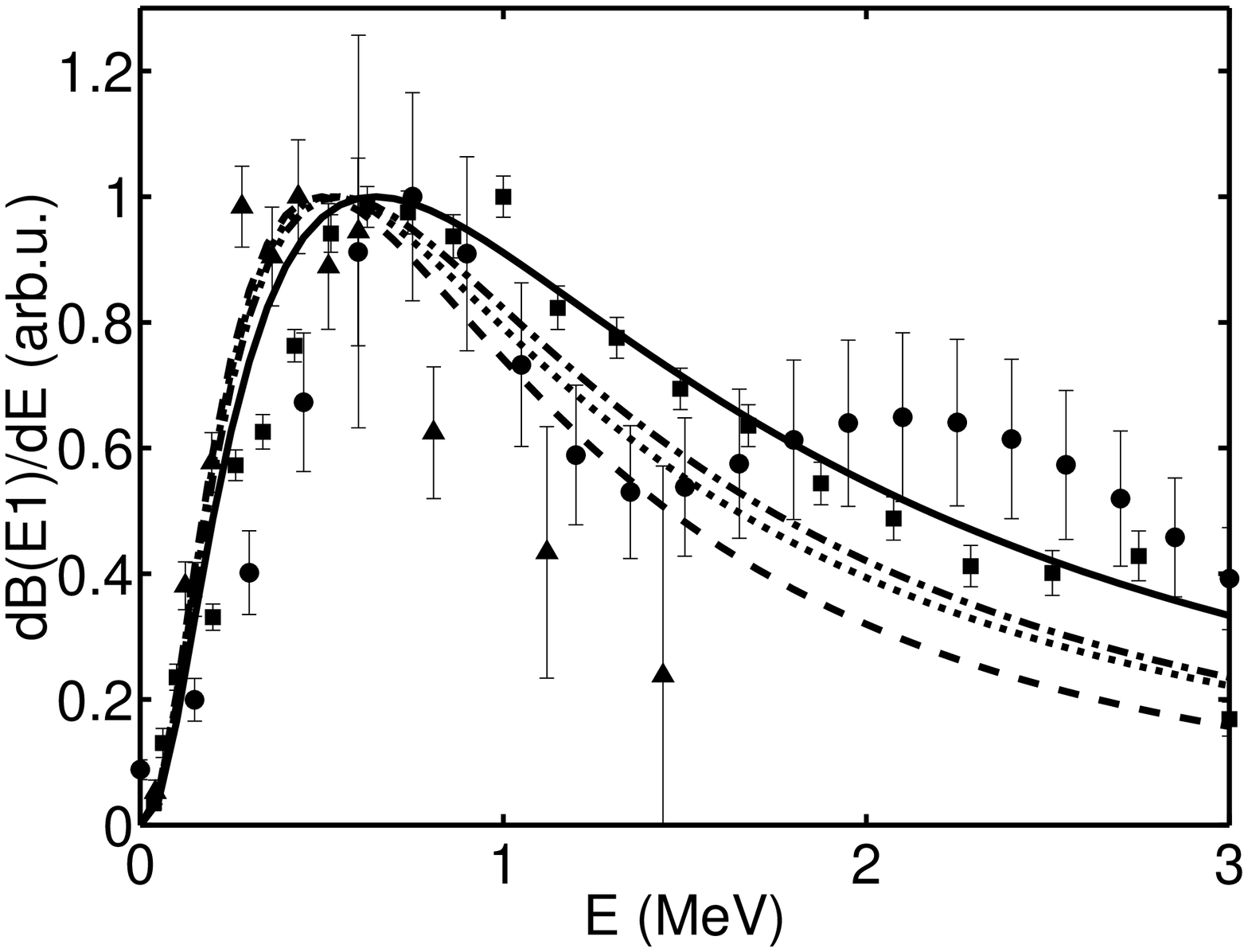}
    \end{minipage}
  \hfill
    \begin{minipage}[t]{0.47\textwidth}
      \centering
      \includegraphics[width=\textwidth]{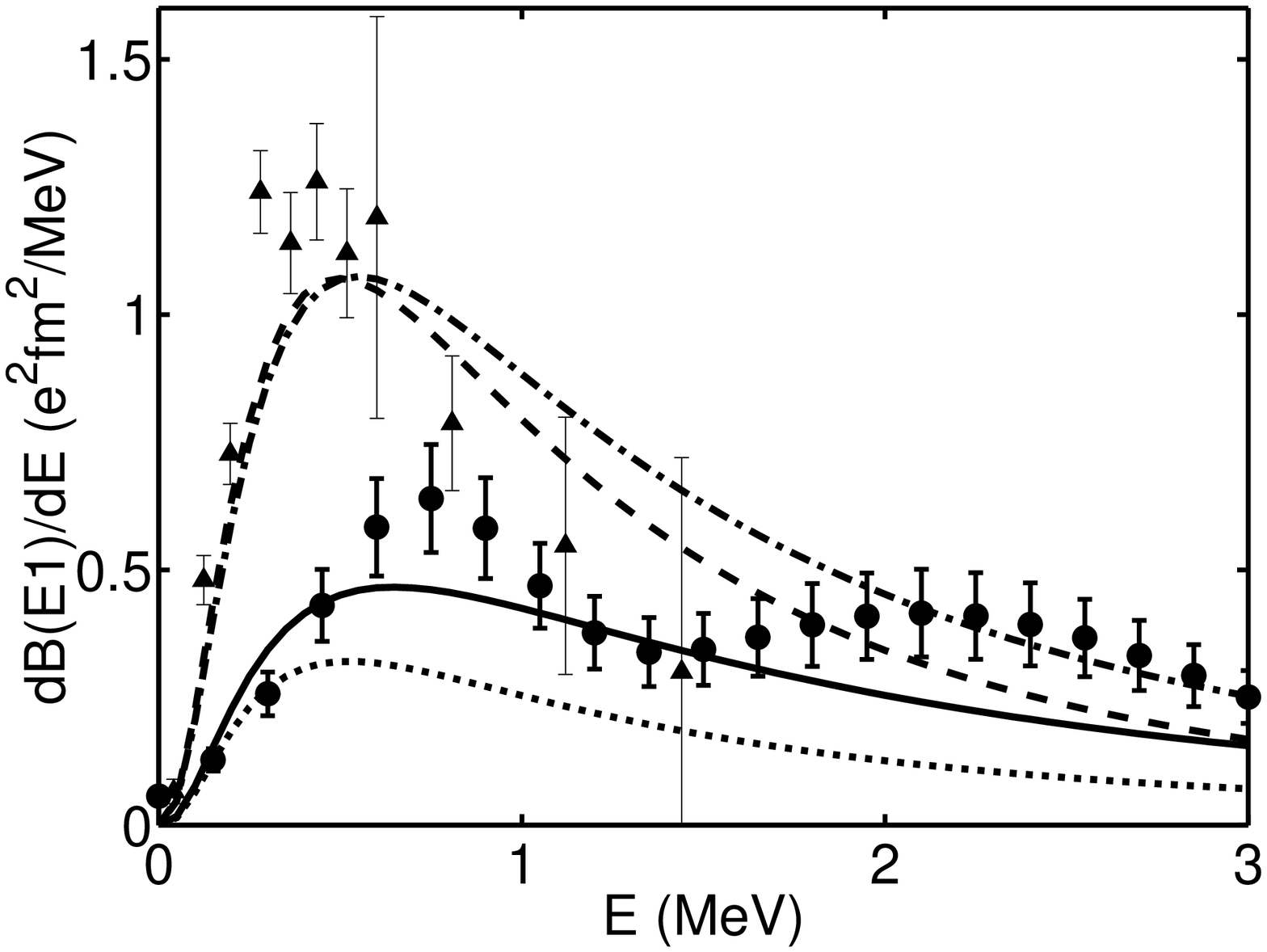}
    \end{minipage}
    \begin{minipage}[b]{0.47\textwidth}
      \centering
      (a)
    \end{minipage}
  \hfill
    \begin{minipage}[b]{0.47\textwidth}
      \centering
      (b)
    \end{minipage}
  \end{minipage}
  \caption{A comparison of our strength function with experimental data
  for \nuc{11}{Li} (circles~\cite{zin97:619}, triangles~\cite{sac93:48},
  squares~\cite{shi95:348}), both in arbitrary units \textnormal{(a)}
  and in absolute scale \textnormal{(b)}. The model WFs are: dotted --
  $\Psi_1$(\nuc{11}{Li}), dashed -- $\Psi_2$(\nuc{11}{Li}), solid --
  $\Psi_3$(\nuc{11}{Li}), dash--dotted -- $\Psi_4$(\nuc{11}{Li}), see
  Table~\ref{tab:wfs}.%
    \label{fig:li11strength}}
  \end{center}
\end{figure}

The results of our calculation are presented in
Fig.~\ref{fig:li11strength} along with available experimental data. The
E1 strength functions are plotted in the figure. In
Fig.~\ref{fig:li11strength}(a) the results are presented in arbitrary
units and in Fig.~\ref{fig:li11strength}(b) they are plotted in absolute
scale.  There exist three published sets of experimental data on EMD
which show large ambiguities. The experiments were performed at
different energies, Zinser~\textit{et~al.}~\cite{zin97:619} used a high
energy beam (280~MeV/A) while Sackett~\textit{et~al.}~\cite{sac93:48}
and Shimoura~\textit{et~al.}~\cite{shi95:348} had much lower energy, 28
and 43~MeV/A, respectively. The validity of extracting the E1 strength
function from the low--energy experiments~\cite{sac93:48,shi95:348} is
discussed in~\cite{sac93:48}. Sackett~\textit{et~al.} report their data
only up to $E=1.45$~MeV above particle--decay threshold, and
Shimoura~\textit{et~al.} list their strength function only in arbitrary
units due to the lack of a forward--angle detector. In the
Zinser~\textit{et~al.} experiment, the strength function was decomposed
into two Gaussian components. The two--peak structure observed by
Zinser~\textit{et~al.} was not seen in the other two experiments. It is
clear that the data sets are contradictory. The shapes are different,
and the peak amplitude is much higher in the Sackett~\textit{et~al.}
data than in the Zinser~\textit{et~al.} data. However, in all data sets
the position of the maximum is approximately the same.

Our results may be commented as follows. All our ground--state WFs give
the peak position at about $0.5-0.7$~MeV which is very close to all
experimental findings. This value is expected.  Indeed, for not too high
energy the predominant contribution to the strength function comes from
large distances, and in our model it is provided by the first, longer
range exponential in Eq.~\eqref{eq:2pwf}, see
also~\cite{for02:697}. Because of this the strength function is mainly a
function of $E/E_0$. Moreover, in the peak region the dominating final
state has $K=1$ which corresponds to the contribution of the single term
with $F_1$ in Eq.~\eqref{eq:str}. Due to these facts the positions of
the peak are close to each other for all our ground--state WFs and
proved to be about $2 E_0$, see also~\cite{pus96:22}. However, terms
with larger $K$ in \eqref{eq:str} change the behaviour at higher
energies.  There is no possibility to explain the two--peak structure of
Zinser~\textit{et~al.} in our model. The one--parameter WF,
$\Psi_1$(\nuc{11}{Li}), leads to a very low strength which is expected
since it underestimates the size of the system, see
Table~\ref{tab:wfs}. The results obtained with our three other model
ground--state WFs of \nuc{11}{Li} are to be compared with experiment if
the absolute values of the strength function are considered.

We see that the $\Psi_2$(\nuc{11}{Li}) WF, which corresponds to
$s$--motion of the valence neutrons with respect to the core and also to
relative $s$--motion between them (or a $(s_{1/2})^2$ configuration in
core--centered, shell--model coordinates), compares rather well with the
Sackett~\textit{et~al.} data. However, for energies $E > 0.5$~MeV the
error bars in~\cite{sac93:48} are large. On the contrary, the
$\Psi_3$(\nuc{11}{Li}) WF, where we have equal weights of $(s_{1/2})^2$
and $(p_{1/2})^2$ configurations in \nuc{11}{Li}, leads to an excellent
agreement with the shape of the Shimoura~\textit{et~al.} data and a
relatively good agreement with the absolute values of the
Zinser~\textit{et~al.}  data. Finally, the $\Psi_4$(\nuc{11}{Li}) WF
leads to results which are significantly different from those obtained
with $\Psi_3$(\nuc{11}{Li}). This may be related to the fact that the
corresponding difference in the relative sign of the $(s_{1/2})^2$ and
$(p_{1/2})^2$ components of the WF leads to a change of the internal
geometry of the system, and the $r_c$ value in particular, as it is seen
from Table~\ref{tab:wfs}. Thus there exists a sensitivity to the
relative sign of the two components in the WF and not only to their
weights. In conclusion we state that this kind of analysis, when applied
to a reliable set of experimental EMD data, would give valuable
information concerning the relative role of the $(s_{1/2})^2$ and
$(p_{1/2})^2$ configurations in the \nuc{11}{Li} ground state.
%
\subsection{\nuc{14}{Be}}
The existing experimental data on \nuc{14}{Be} are scarce. Even such a
fundamental quantity as the binding energy is characterized by large
uncertanties. Combining the two published measurements gives the
tabulated two--neutron separation energy of $1.34 \pm
0.11$~MeV~\cite{gil84:30,wou88:331}. The most recent value of the rms
radius is $3.10 \pm 0.15$~fm~\cite{suz99:658}. Experimental and
theoretical efforts are presently focused on understanding the
ground--state structure. Assuming \nuc{12}{Be} to be a nucleus where the
neutron $p$--shell is closed one gets $0d_{5/2}$ as the next orbital for
the valence neutrons. Several experiments have indeed confirmed a state
around 2~MeV in \nuc{13}{Be} which is assigned having spin--parity
$I^\pi = 5/2^+$~\cite{ost92:343,kor95:343,bel98:636,tho00:63}. However,
in order to reproduce the large size and narrow momentum distributions,
a large $s$--wave component is needed in the ground state of
\nuc{14}{Be}~\cite{des95:52,tho96:53}. The $s$--wave potential fitted
in~\cite{des95:52,tho96:53} to reproduce the \nuc{14}{Be} binding
energy, with the known position of the $d_{5/2}$ state, leads to an
$s$--wave bound state in \nuc{13}{Be} which is clearly not
observed. This problem may be connected with the fact that the
\nuc{12}{Be} nucleus has no closed neutron $p$--shell
(see~\cite{nav00:85,iwa00:481}) and this effect has not been taken into
account in~\cite{tho96:53}. Recent experiments still indicate the
presence of a low--lying virtual $s$--wave state in
\nuc{13}{Be}~\cite{tho00:63}. Furthermore, studies of the isobaric
analog state of \nuc{14}{Be} give an $s$--wave spectroscopic factor of
$45 \pm 20$~\% in the \nuc{14}{Be} ground state~\cite{tak01:515} which
is supported by conclusions made in other
experiments~\cite{suz99:658,lab01:86}. The $K=0$ component of our model
WF provides $s$--motion between all the particles, and we used the $K=4$
component to include some $d$--motion. However, since the $K=4$
component leads to a very broad distribution of strength the shape will
mainly be determined by the $K=0$ component.

\begin{figure}[hbt]
  \begin{center}
  \begin{minipage}{0.95\textwidth}
    \begin{minipage}[t]{0.47\textwidth}
      \centering
      \includegraphics[width=\textwidth]{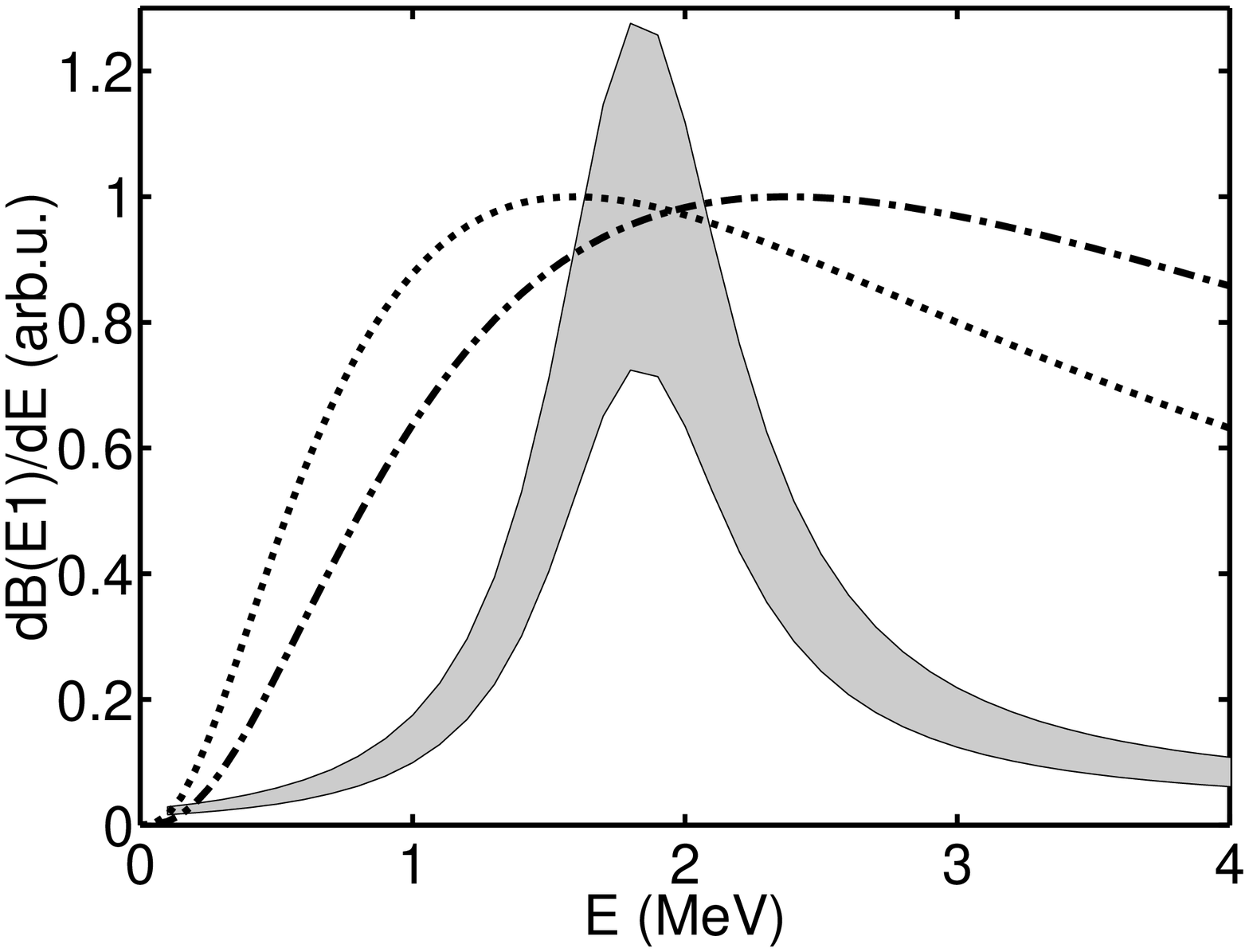}
    \end{minipage}
  \hfill
    \begin{minipage}[t]{0.47\textwidth}
      \centering
      \includegraphics[width=\textwidth]{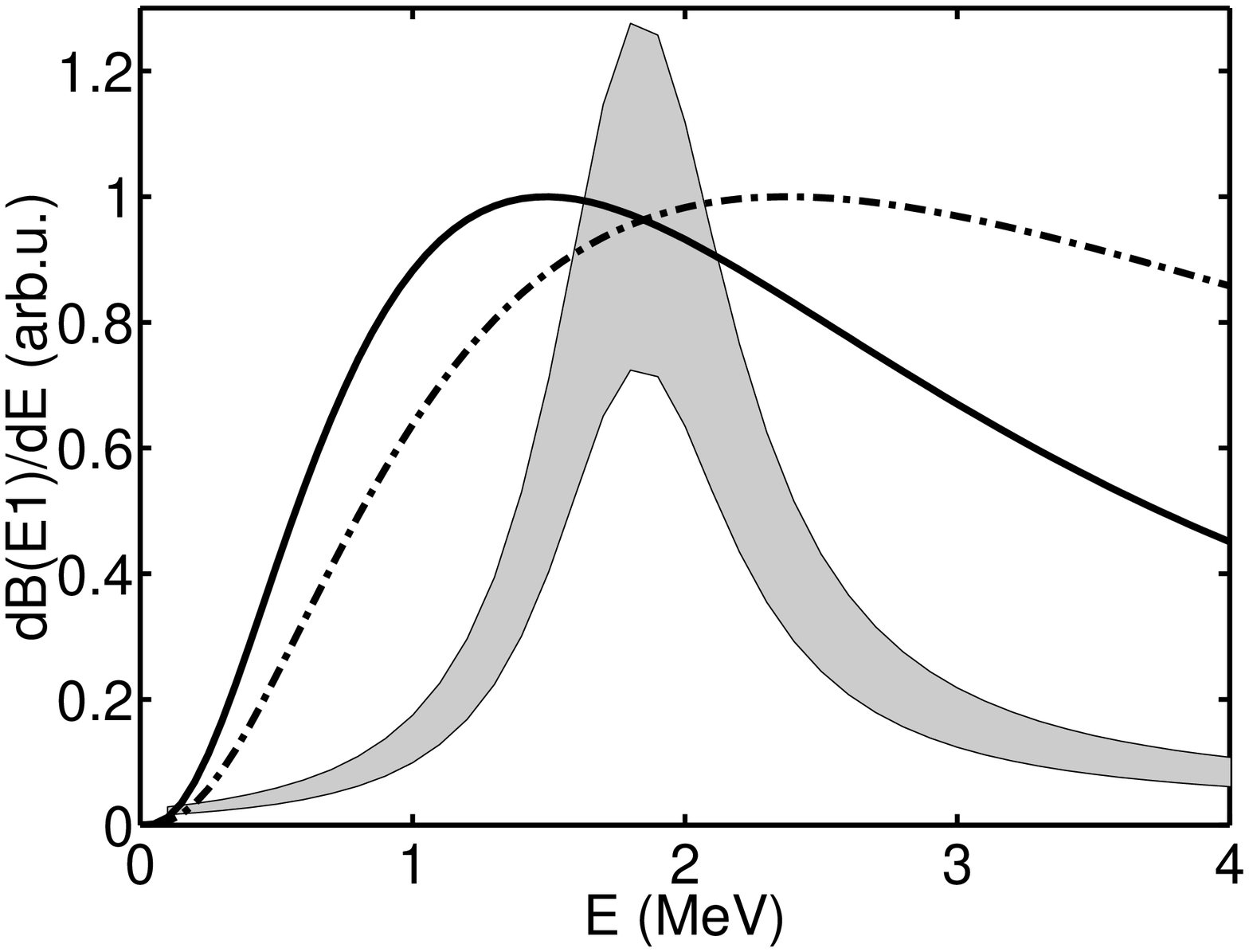}
    \end{minipage}
    \begin{minipage}[b]{0.47\textwidth}
      \centering
      (a)
    \end{minipage}
  \hfill
    \begin{minipage}[b]{0.47\textwidth}
      \centering
      (b)
    \end{minipage}
  \end{minipage}
  \caption{The positions of the peak obtained with two different binding
  energies of \nuc{14}{Be} \textnormal{(a)}. The curves correspond to
  the one--parameter function~\eqref{eq:1pwf} with $E_0 = 1.34$~MeV
  (dash--dotted) and $E_0 = 0.90$~MeV (dotted). In \textnormal{(b)} it
  is shown how the distribution changes when trying to fit both binding
  energy and size. Here the solid curve corresponds to the
  two--parameter $\Psi_3$(\nuc{14}{Be}) WF, Eq.~\eqref{eq:2pwf2}, which
  reproduces both these quantities, while the dash--dotted curve is the
  same as in \textnormal{(a)}. In both figures a comparison is made with
  "experimental data" obtained by dividing out the spectrum of virtual
  photons from the energy spectrum of Ref.~\cite{lab01:86}.%
    \label{fig:be14peak}}
  \end{center}
\end{figure}

The EMD energy spectrum for \nuc{14}{Be} has recently been measured by
Labiche \textit{et~al.}~\cite{lab01:86}. This experiment was performed
at 35~MeV/A. In their analysis the energy spectrum was modeled using the
Breit--Wigner shape with parameters $E_\mathrm{BW} = 1.8 \pm 0.1$~MeV
and $\Gamma_\mathrm{BW} = 0.8 \pm 0.4$~MeV, see Fig.~2
in~\cite{lab01:86}. We prefer to compare theory and experiment using
strength functions which facilitates comparison with EMD cross sections
for other nuclei and at other beam energies as well. We therefore
extracted the dipole strength function from the EMD spectrum of
Ref.~\cite{lab01:86} dividing out the virtual photon spectrum entering
Eq.~\eqref{eq:xsec}. This results in a very narrow distribution peaked
just below 2~MeV as can be seen from Fig.~\ref{fig:be14peak}. Note that
the shape of this strength function is very different from those of
strength functions of other Borromean halo nuclei.  It should also be
mentioned that in the experiment~\cite{lab01:86} only \nuc{12}{Be} and
neutrons were detected. Therefore, we should consider the strength
function up to the \nuc{12}{Be} threshold at 3.2~MeV because above this
energy other reaction channels are open.

In a first step of the analysis we use the position of the peak to check
the binding energy.  If the ground--state WF has an appreciable $K=0$
weight then, as mentioned above, the peak should be positioned at
approximately $2 E_0$. As can be seen from Fig~\ref{fig:be14peak}(a) the
experimental spectrum would correspond to a binding energy of $E_0
\approx 0.9$~MeV, considerably lower than the tabulated value of
1.34~MeV. In fact, we have one more indication that the tabulated
binding energy is too large. Namely, the binding energy seems to be
incompatible with the radius of \nuc{14}{Be}. In Table~\ref{tab:sizes} we
compare the hyperradii
\begin{displaymath}
\rho_\urms = \left \langle \sum_{i=1}^3 A_i(\vec{r}_i-
\vec{R}_\ucm)^2 \right \rangle^ {1/2}
\end{displaymath}
of \nuc{6}{He}, \nuc{11}{Li} and \nuc{14}{Be} which represent the
three--body sizes of the systems.  They are calculated from the rms
radii of these nuclei and their three--body constituents.  For halo
nuclei the hyperradii are mainly determined by the asymptotic behavior
of the WFs. Therefore their values should exhibit a correlation with
binding energies.  Due to this reason one would expect from the
systematics of \nuc{6}{He} and \nuc{11}{Li} that either the binding
energy of \nuc{14}{Be} should be smaller at given size or the size
should be smaller at given energy.

Examples of $K=0$ ground--state WFs that we have used are summarized in
Table~\ref{tab:wfs14be}.  We were unable to reproduce both binding
energy and size with the two--parameter hyperradial
function~\eqref{eq:2pwf}. In the limit $\kappa_1 \rightarrow \kappa_0$
we get $R_\urms = 3.0$~fm. The asymptotic behaviour of the WF changes in
this limit.  In order to reproduce both binding energy and size we have
used one more model hyperradial function
\begin{equation}
  \chi^{(3)} (\rho) \equiv N_0 \left( \frac{\rho}{\rho + \rho_0}
  \right)^{5/2} \exp( -\kappa_0 \rho )
\label{eq:2pwf2}
\end{equation}
which behaves correctly both at large and small $\rho$.  The strength
functions, obtained with the one--parameter function
$\Psi_1$(\nuc{14}{Be}) and with the two--parameter function of
Eq.~\eqref{eq:2pwf2} $\Psi_3$(\nuc{14}{Be}), are compared in
Fig.~\ref{fig:be14peak}(b). Note that the transition ME must be
calculated numerically when \eqref{eq:2pwf2} is used. We see that unlike
the \nuc{6}{He}~\cite{for02:697} and \nuc{11}{Li} cases the low--energy
part of the strength function, and thus the peak position, changes
substantially when going from the one-- to the two--parameter WF. This
happens due to the fact that, at the binding energy and size of
\nuc{14}{Be} used for fitting our bound--state WF, the non--asymptotic
part of the hyperradial function contributes noticeably to the results.

We also tried to vary the binding energy, the radius and the admixture
of different components in the ground--state WF. However in all the
cases we were unable to reproduce the narrow width and large amplitude
of the experimental spectrum. A dominant $K=4$ component in the
ground--state WF leads to a very broad distribution of E1 strength with
a small maximum at high energy (about $5 E_0$). When initial and final
state WFs are obtained from model interparticle potentials, such as the D4
potential of Ref.~\cite{tho96:53}, it is also not possible to reproduce
the shape of the spectrum.  Thus inclusion of FSI does not solve the
problem.  (Note that while the D4 WF reproduces the binding energy and
the size of \nuc{14}{Be} the corresponding potential leads to an unphysical
$s$--wave bound state in \nuc{13}{Be}.)

In Table~\ref{tab:be14strength} we compare the integrated strength
extracted from the experimental data~\cite{lab01:86} with the strengths
obtained with our $\Psi_3$(\nuc{14}{Be}) WF and with the D4 WF of
Ref.~\cite{tho96:53}. When the integration is performed up to the
\nuc{12}{Be} threshold the theoretical strengths prove to be
substantially lower than the experimental one.  Furthermore, an estimate
of the strength integrated up to infinity can be made using the
experimental radius, $R_\urms = 3.10 \pm 0.15$~fm~\cite{suz99:658},
together with the E1 non--energy--weighted cluster sum rule
\begin{equation}
  \int_0^\infty \frac{\d B (\uE 1)}{\d E} \d E =
  \frac{3}{4\pi} Z_\uc^2 e^2 \langle r_\uc^2 \rangle.
\label{eq:newcsr}
\end{equation}
A simple relation between $R_\urms$ and the distance $r_\uc$ between the
CM and the core in a three--body picture is obtained by assuming
$\langle x^2 \rangle = \langle y^2 \rangle = \langle \rho^2 \rangle/2$
(see Appendix A.1 of Ref.~\cite{for02:697}). This gives
\begin{displaymath}
  \langle r_\uc^2 \rangle = \frac{R^2_\urms (A)}{A_\uc} -
  \frac{R^2_\urms (A_\uc)}{A}. 
\end{displaymath}
It is clear that the integrated strength obtained from the EMD
experiment~\cite{lab01:86} is too large as compared with the estimates
in the table. To fit the experimental strength one needs a larger size
of \nuc{14}{Be} while the correlation with binding energy requires a
smaller size. Thus we encounter one more inconsistency.

In conclusion, we state that there are mutual disagreements between
existing experimental data on the binding
energy~\cite{gil84:30,wou88:331}, the radius~\cite{suz99:658}, and the
EMD energy spectrum~\cite{lab01:86} of \nuc{14}{Be}. In our three--body
model we are unable to reproduce these data all at once. This result is
in line with previous microscopic studies of
\nuc{14}{Be}~\cite{des95:52,tho96:53,lab99:60} and in contrast with our
results for other halo nuclei. At the same time we can confirm that a
large $s$--wave component in the \nuc{14}{Be} ground state is required
to obtain an accumulation of E1 strength at low energies.
%
\section{Discussion%
\label{sec:discussion}}
In Ref.~\cite{for02:697} our analytical model was tested on the
relatively well--known \nuc{6}{He} nucleus. In the present paper we have
studied \nuc{11}{Li} and \nuc{14}{Be} for which both experimental data
and theoretical understanding are much poorer. Concerning \nuc{11}{Li}
we conclude that data on E1 strength functions, analyzed in the
framework of our model, can help to resolve the issue on the relative
content of $(s_{1/2})^2$ and $(p_{1/2})^2$ components in the ground
state of \nuc{11}{Li}. However, the existing sets of EMD data are not
consistent with each other. As our calculations show, the E1 strength
function of Sackett \emph{et~al.}~\cite{sac93:48} seems to favor a
configuration with a dominant $(s_{1/2})^2$ component while the data of
Zinser \emph{et~al.}~\cite{zin97:619} and Shimoura
\emph{et~al.}~\cite{shi95:348} are more in agreement with a mixture of
$(s_{1/2})^2$ and $(p_{1/2})^2$ components with equal weights.

Regarding \nuc{14}{Be} we have analyzed the available experimental data
on EMD and found disagreements between these data, the measured size,
and the binding energy.  In the framework of our model it also proved to
be impossible to reproduce the narrow width of the strength
function. This is in contrast with the results for other two--neutron
halo nuclei.  A theoretical study of \nuc{14}{Be} based on model
intercluster potentials led to a similar disagreement~\cite{tho96:53}.
This disagreement is thus not due to the neglection of FSI. The
considerable weight of the $(p_{1/2})^2$ component in \nuc{14}{Be} WF,
suggested in~\cite{lab99:60}, will also not help to reduce the width in
our model. There is a possibility that the structure of \nuc{14}{Be} is
more complicated than for other Borromean, two--neutron halo
nuclei. Perhaps \nuc{14}{Be} has more in common with \nuc{8}{He} than
with \nuc{6}{He} in which case calculations should be done in the
framework of a five--body model or should include excited--core
configurations. More complete experimental data on neutron--removal
cross sections could help to clarify whether \nuc{14}{Be} has a
three--cluster or five--cluster structure.  The five--body analysis of
the experimental data on interaction cross sections, similar to that
performed for \nuc{8}{He}~\cite{alk98:57}, would give a slightly smaller
size of \nuc{14}{Be} than the size obtained in the framework of a
three--body picture~\cite{suz99:658}.  However, in a five--body model
the E1 strength would probably have an even broader distribution still
not allowing the reproduction of the experimental EMD data of Labiche
\emph{et~al.}~\cite{lab01:86}.  Furthermore, methods to measure masses
of short--lived isotopes have improved significantly since the
experiments~\cite{gil84:30,wou88:331} were performed, and new
experimental results on the \nuc{14}{Be} mass would be very important.
\begin{ack}
This work is supported by NFR, Sweden, contract F5102--1484/2001 (C.F.,
M.V.Z.) and RFBR, Russia, grant 00--15--96590 (V.D.E.). V.D.E. also
acknowledges support from the Royal Swedish Academy of Sciences contract
for cooperation between Sweden and former Soviet Union.
\end{ack}
%

%
\newpage
\begin{table}[htp]
    \caption{The model WFs for \nuc{11}{Li}. To reproduce binding energy
    and rms radius we use $\kappa_0 = 0.1193$~fm$^{-1}$ and $\kappa_1 =
    0.3173$~fm$^{-1}$. The quantity $r_\uc$ is the distance between the
    core and the CM. The approximate amplitudes of $(s_{1/2})^2$ and
    $(p_{1/2})^2$ configurations of valence neutrons are shown in the
    last two columns.}
    \vspace{1ex}
    \begin{tabular}{ccccccccc}
      \hline\hline
      WF & $\chi(\rho)$ &
      $a_{00}$ & $a_{20}$ & $a_{21}$ & $R_\urms$ &
      $r_\uc$ & $a_{(s_{1/2})^2}$ & $a_{(p_{1/2})^2}$\\ 
      & Eq. & & & & (fm) & (fm) & &\\
      \hline
      $\Psi_1$(\nuc{11}{Li}) & \eqref{eq:1pwf} & 1 & 0 & 0 & 2.76 
      & 0.60 & $1$ & $0$ \\
      $\Psi_2$(\nuc{11}{Li}) & \eqref{eq:2pwf} & 1 & 0 & 0  
      & 3.55 & 0.95 & $1$ & $0$ \\
      $\Psi_3$(\nuc{11}{Li}) & \eqref{eq:2pwf} & $\sqrt{1/2}$ &
      $-\sqrt{1/6}$ & $\sqrt{1/3}$ & 3.55 & 0.80 & $\sqrt{0.5}$ &
      $\sqrt{0.5}$ \\ 
      $\Psi_4$(\nuc{11}{Li}) & \eqref{eq:2pwf} & $\sqrt{1/2}$ &
      $\sqrt{1/6}$ & $-\sqrt{1/3}$ & 3.55 & 1.08 & $\sqrt{0.5}$ &
      $-\sqrt{0.5}$ \\ 
      \hline\hline 
    \end{tabular}
  \label{tab:wfs}
\vspace{-2.5mm}
\end{table}
\begin{table}[hbt]
  \caption{Comparison of hyperradii and binding energies of three
  Borromean halo nuclei.}
  \vspace{1ex}
  \begin{minipage}{\textwidth}
      \begin{tabular}{c r@{.}l c}
        \hline
        \hline
        Nucleus & \multicolumn{2}{c}{$E_0$~(MeV)} & $\rho_\urms$~(fm) \\
        \hline
        \nuc{11}{Li} & 0&295 & 9.5 \\
        \nuc{6}{He} & 0&97 & 5.4 \\
        \nuc{14}{Be} & 1&34 & 7.4 \\
        \hline
        \hline
      \end{tabular}
  \end{minipage}
  \label{tab:sizes}
\vspace{-2.5mm}
\end{table}
\begin{table}[htp]
    \caption{The model WFs for \nuc{14}{Be}. The binding energy of $E_0 =
    1.34$~MeV corresponds to $\kappa_0 = 0.2543$~fm$^{-1}$ while $E_0 =
    0.90$~MeV corresponds to $\kappa_0 = 0.2084$~fm$^{-1}$. The
    parameter $\rho_0$ entering the $\Psi_3$(\nuc{14}{Be}) WF,
    Eq.~\eqref{eq:2pwf2}, equals 5.42~fm. The quantity $r_\uc$ is the
    distance between the core and the CM. Only the $K = 0$ term in the
    HH expansion has been retained.}
    \vspace{1ex}
    \begin{tabular}{ccccc}
      \hline\hline
      WF & $\chi(\rho)$ & $E_0$ & $R_\urms$ & $r_\uc$ \\ 
      & Eq. & (MeV) & (fm) & (fm) \\
      \hline
      $\Psi_1$(\nuc{14}{Be}) & \eqref{eq:1pwf} & 1.34 & 2.51 & 0.21 \\
      $\Psi_2$(\nuc{14}{Be}) & \eqref{eq:1pwf} & 0.90 & 2.56 & 0.26 \\
      $\Psi_3$(\nuc{14}{Be}) & \eqref{eq:2pwf2}& 1.34 & 3.10 & 0.57 \\
      \hline\hline 
    \end{tabular}
  \label{tab:wfs14be}
\vspace{-2.5mm}
\end{table}
\begin{table}[hbt]
\begin{center}
  \caption{Experimental and theoretical integrated E1 strengths.  The
  first row represents strengths integrated up to 3.2~MeV (\nuc{12}{Be}
  threshold). In the second row total strengths are shown. The total
  strength in the second column is calculated from the value $R_\urms =
  3.10 \pm 0.15$~fm~\cite{suz99:658} using the non--energy--weighted
  cluster sum rule. To this aim, the $r_\uc$ value of $0.57 \pm
  0.07$~fm was obtained from the above value of $R_\urms$ assuming a
  simple three--body structure.}
  \vspace{1ex} 
  \begin{minipage}{\textwidth}
  \begin{tabular}{l c c c c}
    \hline
    \hline
    & Extracted from~\cite{lab01:86} & Obtained from
    $R_\urms$~\cite{suz99:658} &
    $\Psi_3$(\nuc{14}{Be}) & D4~\cite{tho96:53} \\
    \hline
    $B_{E<3.2} (\uE 1)$~($e^2$fm$^2$) & $1.40 \pm 0.40$ & & $0.70$ & 0.96 \\
    $B (\uE 1)$~($e^2$fm$^2$) & & $1.23 \pm 0.33$ & $1.23$ & 1.56 \\
    \hline
    \hline
  \end{tabular}
  \end{minipage}
  \label{tab:be14strength}
\end{center}
\vspace{-2.5mm}
\end{table}
%
\end{document}